\documentstyle[12pt]{article}

\catcode`\@=11
\long\def\@makefntext#1{
\protect\noindent \hbox to 3.2pt {\hskip-.9pt
$^{{\ninerm\@thefnmark}}$\hfil}#1\hfill}		

\def\@makefnmark{\hbox to 0pt{$^{\@thefnmark}$\hss}}  

\def\ps@myheadings{\let\@mkboth\@gobbletwo
\def\@oddhead{\hbox{}
\rightmark\hfil\ninerm\thepage}
\def\@oddfoot{}\def\@evenhead{\ninerm\thepage\hfil
\leftmark\hbox{}}\def\@evenfoot{}
\def\sectionmark##1{}\def\subsectionmark##1{}}

\renewcommand{\thefootnote}{\fnsymbol{footnote}}

\newcounter{sectionc}\newcounter{subsectionc}\newcounter{subsubsectionc}
\renewcommand{\section}[1] {\vspace*{0.6cm}\addtocounter{sectionc}{1}
\setcounter{subsectionc}{0}\setcounter{subsubsectionc}{0}\noindent
	{\normalsize\bf\thesectionc. #1}\par\vspace*{0.4cm}}
\renewcommand{\subsection}[1] {\vspace*{0.6cm}\addtocounter{subsectionc}{1}
	\setcounter{subsubsectionc}{0}\noindent
	{\normalsize\it\thesectionc.\thesubsectionc. #1}\par\vspace*{0.4cm}}
\renewcommand{\subsubsection}[1]
{\vspace*{0.6cm}\addtocounter{subsubsectionc}{1}
	\noindent {\normalsize\rm\thesectionc.\thesubsectionc.\thesubsubsectionc.
	#1}\par\vspace*{0.4cm}}

\newcounter{appendixc}
\newcounter{subappendixc}[appendixc]
\newcounter{subsubappendixc}[subappendixc]

\renewcommand{\appendix}[1] {\vspace*{0.6cm}
        \refstepcounter{appendixc}
        \setcounter{figure}{0}
        \setcounter{table}{0}
        \setcounter{equation}{0}
        \renewcommand{\thefigure}{\Alph{appendixc}.\arabic{figure}}
        \renewcommand{\thetable}{\Alph{appendixc}.\arabic{table}}
        \renewcommand{\theappendixc}{\Alph{appendixc}}
        \renewcommand{\theequation}{\Alph{appendixc}.\arabic{equation}}
        \noindent{\bf Appendix \theappendixc #1}\par\vspace*{0.4cm}}

\def\abstracts#1{{

\centering{\begin{minipage}{12.2truecm}\footnotesize\baselineskip=12pt\noindent
	\centerline{\footnotesize ABSTRACT}\vspace*{0.3cm}
	\parindent=0pt #1
	\end{minipage}}\par}}

\renewenvironment{thebibliography}[1]
	{\begin{list}{\arabic{enumi}.}
	{\usecounter{enumi}\setlength{\parsep}{0pt}
\setlength{\leftmargin 1.25cm}{\rightmargin 0pt}
	 \setlength{\itemsep}{0pt} \settowidth
	{\labelwidth}{#1.}\sloppy}}{\end{list}}

\newcounter{itemlistc}
\newcounter{romanlistc}
\newcounter{alphlistc}
\newcounter{arabiclistc}

\newcommand{\fcaption}[1]{
        \refstepcounter{figure}
        \setbox\@tempboxa = \hbox{\footnotesize Fig.~\thefigure. #1}
        \ifdim \wd\@tempboxa > 6in
           {\begin{center}
        \parbox{6in}{\footnotesize\baselineskip=12pt Fig.~\thefigure. #1}
            \end{center}}
        \else
             {\begin{center}
             {\footnotesize Fig.~\thefigure. #1}
              \end{center}}
        \fi}

\newcommand{\tcaption}[1]{
        \refstepcounter{table}
        \setbox\@tempboxa = \hbox{\footnotesize Table~\thetable. #1}
        \ifdim \wd\@tempboxa > 6in
           {\begin{center}
        \parbox{6in}{\footnotesize\baselineskip=12pt Table~\thetable. #1}
            \end{center}}
        \else
             {\begin{center}
             {\footnotesize Table~\thetable. #1}
              \end{center}}
        \fi}

\def\@citex[#1]#2{\if@filesw\immediate\write\@auxout
	{\string\citation{#2}}\fi
\def\@citea{}\@cite{\@for\@citeb:=#2\do
	{\@citea\def\@citea{,}\@ifundefined
	{b@\@citeb}{{\bf ?}\@warning
	{Citation `\@citeb' on page \thepage \space undefined}}
	{\csname b@\@citeb\endcsname}}}{#1}}

\newif\if@cghi
\def\cite{\@cghitrue\@ifnextchar [{\@tempswatrue
	\@citex}{\@tempswafalse\@citex[]}}
\def\citelow{\@cghifalse\@ifnextchar [{\@tempswatrue
	\@citex}{\@tempswafalse\@citex[]}}
\def\@cite#1#2{{$\null^{#1}$\if@tempswa\typeout
	{IJCGA warning: optional citation argument
	ignored: `#2'} \fi}}
\newcommand{\citeup}{\cite}

 1
 1
 1

\font\ninerm=cmr9

\def\NPB#1#2#3{Nucl.Phys. B{\bf#1} (19#2) #3}
\def\PLB#1#2#3{Phys.Lett. B{\bf#1} (19#2) #3}

\def\PRT#1#2#3{Phys.Rep. {\bf#1} (19#2) #3}

\def\AEF{A.E. Faraggi}

\begin{document}

\newcommand{\st}{\scriptstyle}
\newcommand{\sst}{\scriptscriptstyle}
\newcommand{\mco}{\multicolumn}
\newcommand{\epp}{\epsilon^{\prime}}
\newcommand{\vep}{\varepsilon}
\newcommand{\ra}{\rightarrow}
\newcommand{\ppg}{\pi^+\pi^-\gamma}
\newcommand{\vp}{{\bf p}}
\newcommand{\ko}{K^0}
\newcommand{\kb}{\bar{K^0}}
\newcommand{\al}{\alpha}
\newcommand{\ab}{\bar{\alpha}}
\def\be{\begin{equation}}
\def\ee{\end{equation}}
\def\bea{\begin{eqnarray}}
\def\eea{\end{eqnarray}}
\def\CPbar{\hbox{{\rm CP}\hskip-1.80em{/}}}

\begin{flushright}
\baselineskip=14pt
IASSNS-HEP-94/114{}\\
{hep-ph/9501288}
\end{flushright}

\centerline{\normalsize\bf $Z_2\times Z_2$ orbifold compactification --}
\baselineskip=22pt
\centerline{\normalsize\bf the origin of}
\baselineskip=16pt
\centerline{\normalsize\bf realistic free fermionic models}

\centerline{\footnotesize Alon E. Faraggi}
\baselineskip=13pt
\centerline{\footnotesize\it School of Natural Sciences, Institute for Advanced
Study}
\baselineskip=12pt
\centerline{\footnotesize\it Olden Lane, Princeton, NJ 08540}
\centerline{\footnotesize E-mail:faraggi@sns.ias.edu }

\vspace*{0.9cm}
\abstracts{
All the realistic free fermionic
models utilize a set of basis vectors, the NAHE set, that correspond
to $Z_2\times Z_2$ orbifold compactification with nontrivial background fields.
I argue that the realistic features of free fermionic models,
like the number of generations and the fermion mass spectrum are due to the
underlying $Z_2\times Z_2$ orbifold compactification.
}

\footnotetext{To appear in Proceedings of the Beyond the Standard Model IV
Conference, Lake Tahoe, California, December 13--18, 1994.}

\normalsize\baselineskip=15pt
\setcounter{footnote}{0}
\renewcommand{\thefootnote}{\alph{footnote}}
\vspace*{0.6cm}

As a unified theory of gravity and the gauge interactions, heterotic
string theory\citeup{heterotic}
should reproduce the matter and symmetry content of the Standard Model
and determine the fermion mass spectrum.
Presently we do not know what is the dynamical mechanism that selects
the unique string vacuum, and, a priori, there is a large number of potentially
viable superstring models.

The notion, however, that there is a huge number of string models is
somewhat misleading. By just imposing one or two phenomenological
criteria, like three generations and a gauge group that can be reduced
at low energies to the standard model gauge group, already one finds
that the number of possibilities is substantially reduced. Imposing
further phenomenological constraints may indeed single out a unique
superstring model. If such a model is constructed, it will certainly be
of use in trying to learn about the dynamical mechanism that
chooses the string vacuum.

The task of constructing phenomenologically viable string models
seems hopeless. While in ten dimensions the string vacuum is more or less
unique, in four dimensions there is a huge number of equivalent candidates.
The string consistency constraints impose a number of degrees
of freedom and those degrees of freedom produce a symmetry that is
larger than the observed symmetry at low energies. Furthermore
the number of chiral generations is also determined in the four dimensional
vacuum and is correlated with the gauge degrees of freedom.
A bottom--up approach, in which different blocks of the standard
model are assembled together piece by piece, is not adequate.
Rather, what is required is a top--bottom approach in which the features
of the standard model are carved out of the more symmetric string vacua.

Is there a guiding principle that may distinguish among the equivalent
string vacuum ? String vacua exhibits a new kind
of symmetry, usually referred to as target--space duality\citeup{GPR},
which is a generalization of the $R\rightarrow{1/ R}$
duality in the case of $S^1$.
At the self--dual point, $R_j={1/ R_j}$,
space--time symmetries are enhanced. For appropriate choices of the
background fields the space--time symmetries are maximally
enhanced\citeup{FOC}. At the maximally symmetric point the internal
degrees of freedom that are needed to cancel the conformal anomaly
may be represented in terms of internal free fermions propagating on
the string world--sheet. It is plausible that if string theory is realized
in nature then the  true string vacuum is in the vicinity of the highly
symmetric self--dual point. It may turn out that
near that point the free fermionic formulation\citeup{FFF} provides
a good approximation to the true string vacuum. However,
the number of consistent free fermionic models is still enormous.

As is well known in (2,2) string models that admit a geometrical
interpretation the number of chiral generations is half
the Euler number of the six dimensional compactified manifold.
Following LEP data it is plausible to assume that only three
complete generations exist in nature. How can three generations
arise from a six dimensional compactified space. The answer
may be simple. The six dimensional compactified
space is divided into three factors of two.
In the orbifold language\citeup{DHVW},
divide the six dimensional space,
which is compactified on a flat torus, by a $Z_2\times Z_2$
discrete symmetry. In that case the $Z_2\times Z_2$ orbifold
model produces exactly three twisted sectors.
In the $Z_2\times Z_2$ orbifold
on a six dimensional space, the number three is deeply rooted in the
structure of the models. Thus, the $Z_2\times Z_2$ can very naturally
lead to models with three generations. Namely, each light
generation comes from a different twisted sector of the
$Z_2\times Z_2$ orbifold model.

It appears that $Z_2\times Z_2$ orbifold on the flat torus
of the six dimensional compactified space, can very naturally lead to
three generations. However, in general, $Z_2\times Z_2$ orbifold
on generic lattices do not lead to three generation models.
For example the $Z_2\times Z_2$ orbifold on $SO(4)^3$ lattice
did not yield three generation models.
In contrast, the $Z_2\times Z_2$ models at the free fermionic point
in toroidal compactification space,
realized by the NAHE set\citeup{REVAMP,SLM}, do produce
three generation models. The difference is seen by examining the
number of fixed points in the two compactifications with (2,2)
world--sheet supersymmetry. On the $SO(4)^3$ lattice the $Z_2\times Z_2$
produces sixteen generations,
from each twisted sector.  On the
$SO(12)$ lattice, which corresponds to the free fermionic point in
the toroidal compactification space, it produces
eight chiral generations, from each twisted
sector. In the fermionic three generation constructions each one
of the complex planes of the $Z_2\times Z_2$ orbifold is
modded out by additional $Z_2^3$ symmetries, thus reducing the number
of generations to one generation from each twisted sector.

In the (2,2) fermionic constructions one starts from a set of
boundary condition vectors that produces an $N=4$ supersymmetric model
with $SO(12)\times E_8\times E_8$ gauge group\citeup{FOC}. One then
adds two boundary condition vectors that correspond to the $Z_2\times Z_2$
twisting. The resulting gauge group is $SO(4)^3\times E_6\times U(1)^2\times
E_8$ with $N=1$ space--time supersymmetry.
In this model there are twenty four chiral generations from the
boundary condition vectors that correspond to twisted sectors and
three pairs of chiral and anti chiral generations from the untwisted sector.
The number of twisted and untwisted moduli is equal to the number of
generations. In addition the untwisted and twisted sectors produce
$E_6\times E_8$ singlets
that are obtained by acting on the vacuum with oscillators that arise from the
fermionic degrees of freedom that correspond to the six internal compactified
dimensions.

In the orbifold formulation\citeup{DHVW}
the same model is obtained by applying a
$Z_2\times Z_2$ twist to a torodialy compactified $SO(12)$ lattice and
$E_8\times E_8$ gauge symmetry.
The 36 free parameters of the six dimensional
metric and the antisymmetric tensor field parameterize
the six dimensional compactified space.
For generic values of these parameters the gauge symmetry that arises
from the six dimensional compactified torus is $U(1)^6$. For specific
choices of the background parameters
the $U(1)^6$ of the compactified torus is enlarged. To reproduce the
$SO(12)\times E_8\times E_8$ gauge group of
the free fermionic model, the metric
$G_{ij}$ is the Cartan matrix of $SO(12)$ and the antisymmetric tensor
field is given by, $B_{ij}=G_{ij}$ for $i>j$; $B_{ij}=0$ for $i=j$ and
$B_{ij}=-G_{ij}$ for $i<j$.
For $R_I=\sqrt2$ and with the chosen background fields,
the right--moving momenta produce the root vectors of $SO(12)$,
thus reproducing the same gauge group as in the free fermionic model.
The orbifold model is obtained by
moding out the six dimensional torus by a discrete symmetry group.
The massless spectrum contains states from the untwisted and twisted sectors.
In the case of ``standard embedding'' the number of chiral families is given
by one half of the Euler characteristic.
To translate the fermionic boundary conditions to twists and shifts in the
bosonic formulation the real fermionic degrees of freedom that correspond
to the compactified dimensions are bosonized. The fermionic boundary
condition vectors, $b_1$ and $b_2$, then translate to $Z_2\times Z_2$
twist on the compactified coordinates and to shifts on the gauge degrees of
freedom. It is then seen that symmetries and spectrum of the orbifold model
coincide with those of the corresponding fermionic model\citeup{FOC}.

The realistic
free fermionic models correspond to models with (2,0), rather than (2,2),
world--sheet supersymmetry. The transition from the (2,2) models to the
(2,0) models can be regarded as choosing a GSO phase between the two
boundary condition vectors that produce the spinorial of $SO(16)$.
The GSO projection projects out the massless states from these
sectors and the resulting gauge group is $SO(12)\times SO(16)\times SO(16)$,
with $N=4$ space--time supersymmetry.
Alternatively, one of the spinorial vectors may be enlarged with additional
four periodic complex fermions in the hidden sector. The $E_8\times E_8$
gauge group is modified
to $SO(16)\times SO(16)$, as in the first construction.
The analysis with respect to the number of fixed points is identical to the
case with (2,2) world--sheet supersymmetry. However, in this case the
observable gauge group after applying the $Z_2\times Z_2$ is
$SO(10)\times U(1)$ rather than $E_6$, and the $U(1)$ is ``anomalous''.
The twisted sectors produce spinorial and vectorial sixteen of the
observable $SO(10)$ and hidden $SO(16)$ gauge groups, respectively.

The structure of the $Z_2\times Z_2$ orbifold with (2,0) world--sheet
supersymmetry and standard embedding, is common to all the realistic
free fermionic models. Three generation models are obtained by adding
three additional boundary condition basis vectors, beyond the NAHE set.
The additional boundary condition vectors mod out each of the three
complex planes of the $Z_2\times Z_2$ orbifold by a $Z_2^3$ symmetry and
break the observable $SO(10)$ symmetry to one of its maximal
subgroups $SU(5)\times U(1)$, $SO(6)\times SO(4)$ or
$SU(3)\times SU(2)\times U(1)^2$.

The fermion mass spectrum is also seen to originate from the $Z_2\times Z_2$
orbifold structure, realized by the NAHE set. The untwisted sector
produces three pairs of Higgs doublets and a combination of the vectors
that break the $SO(10)$ symmetry produces one or two  additional pairs.
Due to the horizontal symmetries in the $Z_2\times Z_2$ orbifold models,
each pair of Higgs doublets couples only to states from one of the twisted
sectors, producing couplings $16_j16_j10_j$ $j=1,2,3$. The cancelation
of the anomalous $U(1)$ D--term equation by singlet VEVs, gives Planck scale
mass to several Higgs doublets. As a result, there exist models in which
only one mass term, namely the top quark mass term, exist at the cubic
level of the superpotential. The mass terms for the lighter quarks and leptons
are obtained from nonrenormalizable terms. The nonrenormalizable terms
contain $SO(10)$ singlets with nonvanishing VEVs, that are required to
cancel the anomalous $U(1)$ D--term equation. Thus, the
nonrenormalizable terms become effective renormalizable terms that are
suppressed relative to the leading cubic level terms. Due to the horizontal
symmetries and the singlet VEVs one generation is necessarily
light\citeup{NRT}.
Similarly, the mixing terms arise generically from nonrenormalizable
terms of the form $16_i16_j1016_i16_j\phi^n$,
where the first two 16 are in the spinorial representation
of the observable $SO(10)$, the 10 is in the vector representation
of the observable $SO(10)$, the last two 16 are in the vector representation of
the hidden $SO(16)$ and $\phi^n$ is a combination of $SO(10)\times SO(16)$
scalar singlets\citeup{CKM,FOC}. The $Z_2\times Z_2$ orbifold structure
gives rise to the horizontal symmetries that may be needed to understand the
matter mass spectrum. Requiring adequate generation mixing and the form
of the mixing terms necessitates that we give nonvanishing VEVs to some
of the hidden sector 16 representations. In Ref. [10] it was shown
that this is possibly the source of supersymmetry breaking in these models.

{\bf{Acknowledgements}}

This work is supported by DOE grant DE-FG02-90ER40542.

\small
\baselineskip=10pt
\medskip
{\bf{References}}

\end{document}